\documentclass[reprint,aps,pra,superscriptaddress,amsmath,amssymb
]{revtex4-1}

\usepackage[utf8]{inputenc}
\usepackage{graphicx}
\usepackage{dcolumn}
\usepackage{bm}
\usepackage{mathrsfs}
\usepackage{amsfonts}
\usepackage{color}
\usepackage{xcolor}
\usepackage{braket}
\usepackage[english]{babel}
\usepackage{lipsum} 
\usepackage{bbm}
\addto\captionsenglish{}
\usepackage{url}
\definecolor{darkblue}{rgb}{0,0,0.5}
\usepackage[colorlinks=true, allcolors=blue]{hyperref}
\DeclareMathOperator{\sech}{sech}
\DeclareMathOperator*{\argmax}{\arg\!\max}
\DeclareMathOperator{\Tr}{Tr}

\begin{document}

\title{Optical Gottesman-Kitaev-Preskill qubit generation via approximate squeezed coherent state superposition breeding}

\author{Andrew J. Pizzimenti}
\email{ajpizzimenti@arizona.edu}
\affiliation{Wyant College of Optical Sciences, The University of Arizona, 1630 East University Boulevard, Tucson, Arizona 85721, USA}
\author{Daniel Soh}
\affiliation{Wyant College of Optical Sciences, The University of Arizona, 1630 E. University Blvd., Tucson, AZ 85721, USA}

\date{\today}

\begin{abstract}
Gottesman-Kitaev-Preskill (GKP) qubits, known for their exceptional error-correction capabilities, are highly coveted in quantum computing. However, generating optical GKP qubits has been a significant challenge. Measurement-based methods, where a portion of entangled squeezed vacuum modes are measured with photon number resolving detectors heralding a desired state in the undetected modes, have emerged as leading candidates for optical GKP qubit generation due their minimal resource requirements. While the current measurement-based methods can produce high-quality GKP qubits, they suffer from low success probabilities limiting experimental realization. The heart of the problem lies in the duality of photon number resolving measurements, being both the source of nonlinearity needed to generate quality GKP qubits and the component driving down their probability of successful production. Our method, breeding approximate squeezed coherent state superpositions created by generalized photon subtraction, overcomes this problem by supplementing two photon number resolving measurements with a single high success probability homodyne measurement. This scheme achieves success probabilities $\geq 10^{-5}$, two orders of magnitude higher than strictly photon number resolving measurement-based methods, while still producing states with high fidelity, possessing error-correction capabilities equivalent to up to a 10 dB squeezed GKP qubit. This breakthrough significantly advances the practical use of the optical GKP qubit encoding.  
\end{abstract}

\maketitle

\section{Introduction}\label{sec:Introduction}
 Bosonic qubit encodings come in two forms: discrete and continuous variable. In the discrete-variable (DV) encoding, a qubit is most commonly realized by the presence or absence of a single energy quantum excitation. Qubits in the continuous-variable (CV) encoding use two orthogonal quantum states, each of which is typically an infinite superposition of different numbers of energy quantum excitations. One specific CV encoding is the Gottesman-Kitaev-Preskill (GKP) qubit \cite{Gottesman2001}. Being a simultaneous eigenstate of commuting displacement operators that act in nonparallel directions, the GKP qubit has a unique gridlike structure in phase space and is resilient to displacement and excitation loss errors \cite{glancy2006error,Albert2018performance}. GKP qubits in vibrational modes of trapped ions and superconducting microwave cavities have been realized with relative ease due to the large nonlinearities intrinsic to these systems \cite{fluhmann2019encoding,campagne2020quantum}. However, the nonlinearities that allow for straightforward GKP qubit generation in these physical implementations hinder the performance of the Gaussian logical operations on the GKP qubits. Moreover, these material GKP states have significant difficulty in realizing the two-qubit gates necessary for universal quantum computation \cite{DiVincenzo_2000} due to the lack of a natural coupling between qubits in these physical systems \cite{Gao_2018,toyoda2015hong}. In the optical domain, on the contrary, single- and two-qubit gates can all be realized with readily available Gaussian operations \cite{Menicucci_2006}. Scalable architectures for photonic measurement-based quantum computing (MBQC) have already been demonstrated \cite{Larsen_2021, Asavanant_2021}, and the addition of a optical GKP qubit source will make it universal and fault tolerant \cite{Baragiola_2019, Yamasaki_2020}. Despite this strong motivation, development of a high-rate source of quality optical GKP qubits has proven extremely challenging and has held back progress in realizing optical quantum computers.
 
 A promising advancement in optical GKP qubit generation has been the use of single-photon and photon number resolving (PNR) measurements to introduce the nonlinearity needed to generate these exotic states. Gaussian-input measurement-based methods, as opposed to measurement-based methods that employ non-Gaussian states at the input \cite{Eaton_2019}, benefit from experimental simplicity since their single-mode squeezed vacuum (SV) state inputs are processed with passive linear optics and all nonlinearity is confined to the measurements \cite{Tzitrin2020,Fukui2022Efficient}. This simplicity comes at a price: each PNR measurement succeeds probabilistically, meaning overall success probability decreases with each additional measurement. Current implementations have been able to produce high-quality GKP qubits at mediocre rates, with $\sim 10^{-7}$ being the best success probabilities reported for GKP qubits with $\sim 10~ \text{dB}$ of squeezing generated using three or four PNR measurements \cite{takase2023gottesman}. This prompts a question --- can GKP qubits be generated using fewer PNR measurements?
 
 One of the most instructive methods for optical GKP qubit generation is the squeezed coherent state superposition (SCSS) breeding protocol \cite{vasconcelos2010}. Proposed using linear optics and postselected homodyne measurements, the protocol was shown to work with feed-forward measurements and classical postprocessing making the protocol quasideterministic \cite{weigand2018}. The bottleneck of the protocol is the large deeply squeezed SCSSs that are required. A recent advancement towards feasible experimental generation of these states is generalized photon subtraction (GPS) \cite{takase2021generation}. An extension of traditional photon subtraction \cite{Dakna1997generating} and a subcase of Gaussian boson sampling (GBS)-based methods \cite{Hamilton2017,Su_2019,Gagatsos2019,pizzimenti2021}, GPS can generate large deeply squeezed approximate SCSSs at high rates with a single PNR measurement. Here we establish a rigorous performance analysis of a practical implementation of GKP qubit generation via breeding by using GPS states as inputs to the breeding protocol with and without post selection. In doing so, we demonstrate that with two PNR measurements one can generate high-quality GKP qubits up to $10~ \text{dB}$ of squeezing and answering affirmatively to the question posed in the previous paragraph. This was achieved by supplementing the PNR measurements with the high success probability homodyne measurement present in the breeding process. The reduced measurement resources of our method directly translate into a substantially better overall GKP qubit generation probability, i.e., greater than $10^{-5}$, which is two orders of magnitude better than strictly PNR measurement-based methods, when breeding without postselection. In this article, as our main contribution, we provide a rigorous proof that switching from the PNR measurement to homodyne measurement significantly improves the generation probability.
 
The elementary resource requirements of our GPS state breeding (SV and coherent light sources, passive Gaussian optics, PNR and homodyne measurements, classical feed-forward processing) and its superior success rates make it a practically viable route to generating optical GKP qubits experimentally. Indeed, with reasonable assumptions, our calculation reveals that GKP qubits with 8 dB of squeezing could be generated at a MHz rate using just a few hundred of our sources in parallel and current state-of-the-art PNR detectors, such as in \cite{Mattioli2016}. With mild upgrades to the resolution of photon detectors and 1000 of our sources, MHz generation of 10 dB squeezed GKP qubits is feasible. This marks a significant advancement towards optical quantum computing as this is the final element needed to achieve fault tolerance and universality in the MBQC architecture. 

\section{Approximate State Breeding} \label{sec:ApproximateStateBreeding}
Ideal SCSS breeding has been well studied in \cite{vasconcelos2010,Terhal2016}. Instead, we investigate the practical and nonideal implementation of these breeding protocols by deriving the breeding output state wave function, homodyne measurement probability distribution, and corrective displacement amplitude when the inputs are the GPS states.

To begin, we determine the optimal squeezing for the GPS output state to be used in the breeding protocol. The GPS output state wave function, $\psi^{\text{\tiny{GPS}}}_{r}(q|n)$, has peaks at $q = \pm \sqrt{n \sech(2 r)}$, while the SCSSs used for breeding in Refs. \cite{vasconcelos2010,weigand2018} had peaks at $q = \pm \sqrt{2 \pi}$. Therefore, the optimal GPS state to generate GKP qubits via breeding is one with squeezing,
\begin{align}
    r_{\text{opt}}(n) = \frac{1}{2} \sech^{-1}\left(\frac{2 \pi}{n}\right).
    \label{eq:SpacingCondition}
\end{align}
The conditional wave function that results from breeding with $ \psi^{\text{\tiny{GPS}}}_{r_{\text{\tiny{opt}}}}(q|n)$ and its associated homodyne measurement probability distribution are found to be (see Appendix \ref{app:phi_n} for full derivation)
\begin{align}
    \phi_{n}(q_1|p) &  = \frac{e^{-\frac{1}{2}\frac{n}{2 \pi} q_1^2 }}{\mathcal{N}} \sum_{k = 0}^n ~\Theta_k\left(p,\frac{n}{2 \pi}\right) ~(-1)^k \binom{n}{k}q_1^{2(n-k)},
   \label{eq:GPSBreedState}\\
   P_n^{\text{\tiny{Hom}}}(p) & = \left[\frac{1}{  \Gamma\left(\frac{1}{2}+n\right)} \left(\frac{n}{4 \pi}\right)^n \frac{\sqrt{n}}{2 \pi} \right]^2 \times \mathcal{N}^2.     \label{eq:GPSBreedingHomodyneDist}
\end{align}
Here, $\mathcal{N}$ and $\Theta_k$ are given as
\begin{widetext}
    \begin{align}
        \mathcal{N} & = \sqrt{\sum_{k,k'=0}^n ~\Theta_k\left(p,\frac{n}{2 \pi}\right) ~\Theta_{k'}\left(p,\frac{n}{2 \pi}\right)  (-1)^{k+k'} \binom{n}{k} \binom{n}{k'} \left(\frac{2 \pi}{n}\right)^{2n-k-k'+\frac{1}{2}} \Gamma\left( 2n-k-k'+\frac{1}{2} \right)}.
    \label{eq:tilde_phi_norm}
    \end{align}
\end{widetext}
and
\begin{align}
    \Theta_k(y,\gamma) &  = e^{-\frac{y^2}{2 \gamma}} \sum_{l = 0}^{2 k} f^k_l(\gamma)~ y^{2(k-l)} \label{eq:Theta_k},
\end{align}
where
\begin{align}
        f^k_l(\gamma) & =  \left( -\frac{1}{\gamma^2} \right)^{k-l} \binom{2 k}{2 l} \left( \frac{2}{\gamma} \right)^{l+\frac{1}{2}} \Gamma\left( l + \frac{1}{2} \right)\label{eq:fFunction}.
\end{align}
\begin{figure}
    \centering
    \includegraphics[width=0.9\linewidth]{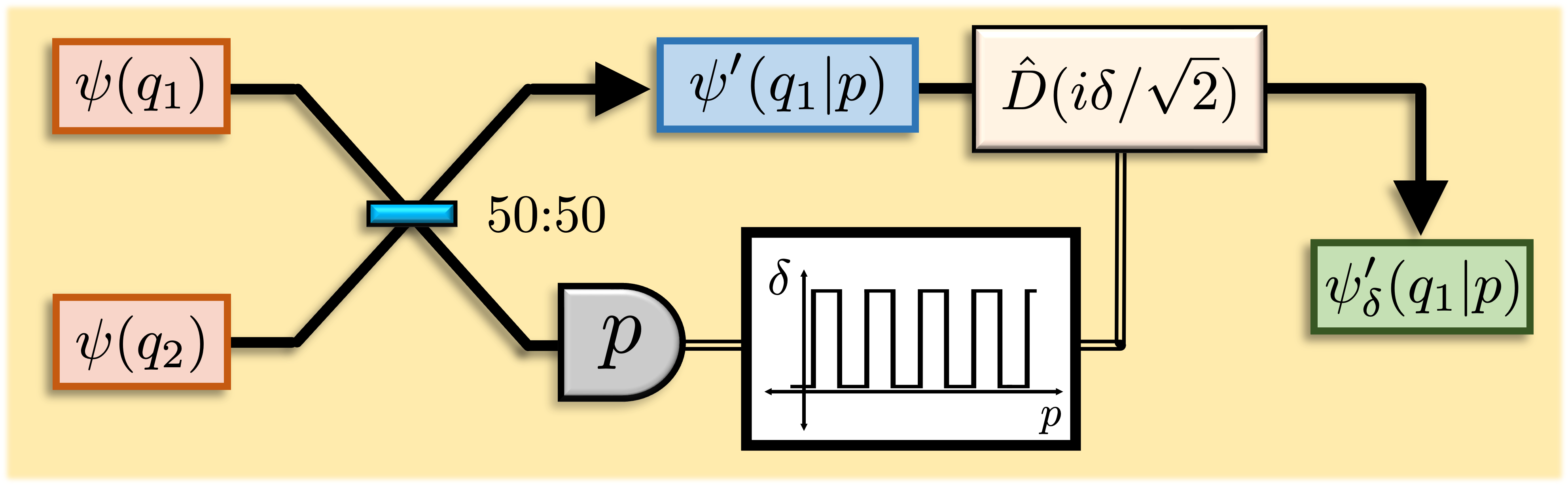}
    \caption{Optical circuit that performs breeding without postselection on two states, $\psi$. The homodyne measurement outcome, $p$, which heralds the state $\psi'$, is classically processed to determine the proper value of $\delta$ for the corrective displacement. The value of $\delta$ is fed forward and $\hat{D}(i \delta / \sqrt{2})$ is applied to $\psi'$ to get the final output state $\psi'_{\delta}(q_1|p)$.}
    \label{fig:CorrectiveDisplacment}
\end{figure}
To perform breeding without postselection, a corrective displacement $\hat{D}( i \delta / \sqrt{2})$ needs to be applied to the breeding output state; see Fig. \ref{fig:CorrectiveDisplacment}. The choice of corrective displacement amplitude amounts to a binary decision based on the homodyne measurement outcome,
\begin{align}
    \delta & =  \begin{cases}
         0, &  g_n(p) \geq 0 \\
         \frac{\sqrt{\pi}}{2}, &  g_n(p) < 0 \\
    \end{cases},
    \label{eq:OptCorrectDisp}
\end{align}
with 
\begin{align}
    g_n(p) & =  \frac{1}{2 \pi \mathcal{N}} \sum_{k,k'=0}^{n} \sum_{j = 0}^{2(n-k)} \sum_{j' = 0}^{2(n-k')}  (-1)^{k+k'} \binom{n}{k} \binom{n}{k'} \nonumber \\
    & \times \Theta_{k}\left(p,\frac{n}{2 \pi}\right) \Theta_{k'}\left(p,\frac{n}{2 \pi}\right) \Theta_{2n-\Delta k - \Delta j}\left(2 \sqrt{\pi}, \frac{4 \pi}{n}\right) \nonumber 
    \\ & \times f_{j}^{n-k}\left(\frac{n}{2 \pi}\right) ~f_{j'}^{n-k'}\left(\frac{n}{2 \pi}\right), 
    \label{eq:ApproxBreedingGFunc}
\end{align}
and $\Delta k = k - k'$, $\Delta j = j - j'$. The state resulting from displacing $\phi_{n}(q_1|p)$ by $\hat{D}(i \beta / \sqrt{2})$ is 
\begin{align}
    \phi_{n,\beta}(q_1|p) & =  \frac{e^{-\frac{1}{2}\frac{n}{2 \pi} q_1^2  -i q_1 \beta}}{2 \pi \mathcal{N}} \sum_{k = 0}^n \sum_{j=0}^{2(n-k)} \sum_{m=0}^{2(n-k-j)} (-1)^{k} \binom{n}{k} \nonumber \\ 
    & \times f^{n-k}_j\left(\frac{n}{2 \pi}\right) ~ f^{n-k-j}_m\left(\frac{2 \pi}{n}\right) \nonumber \\
    & \times \Theta_k\left(p,\frac{n}{2 \pi}\right) q_1^{2(n-k-j-m)}.
    \label{eq:TildePhiDisplaced}
\end{align}
A full derivation of Eqs. \eqref{eq:OptCorrectDisp}, \eqref{eq:ApproxBreedingGFunc}, and \eqref{eq:TildePhiDisplaced} can be found in Appendix \ref{app:CorrectiveDispacmentAndOutputState}. We show, in Appendix \ref{app:ExactStateBreeding}, that approximate state breeding approaches exact state breeding as $r_{\text{\tiny{opt}}}$ increases, as one would expect given the high fidelity between the initial states.

 \begin{figure*}
    \centering
    \includegraphics[width=0.55\paperwidth]{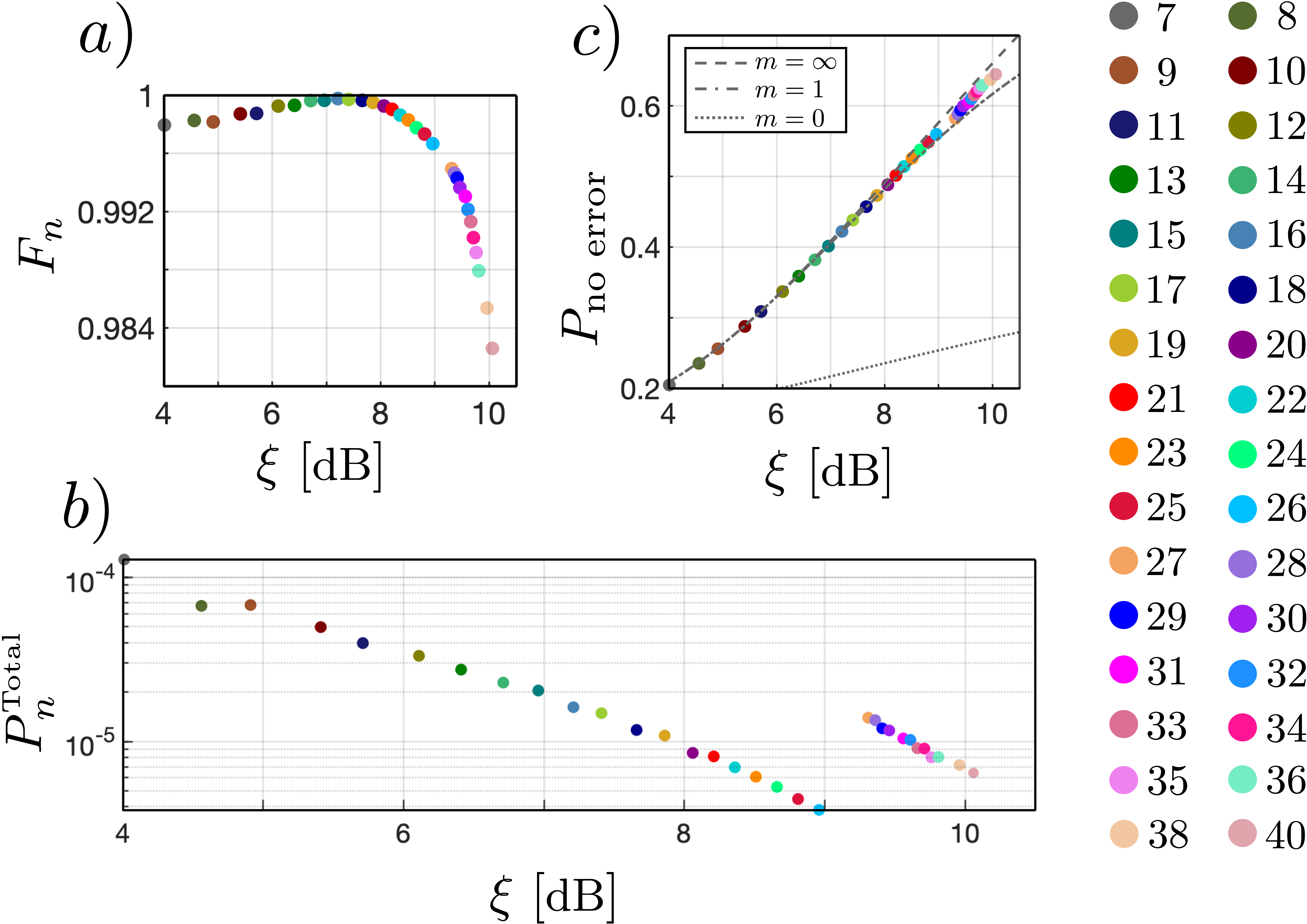}
    \caption{(a) Fidelity of the states found by the optimization in Eq.\eqref{eq:Fargmax} to finite-squeezed logical-zero GKP qubits with squeezing parameter $\xi$. (b) Probability of generating the states with fidelity given in (a). (c) Glancy-Knill parameter associated with the states found by the optimization in Eq. \eqref{eq:Fargmax}. The curves are the Glancy-Knill parameter associated with finite-squeezed logical-zero GKP qubits with squeezing parameter $\xi$, where the summation limits of the finite-squeezed GKP qubit in Ref. \cite{GKPQubit} are taken to be $t=-m$ to $m$. In all plots, the color of each point denotes the number of photons, $n$, associated with the GPS states used to create the states.}
    \label{fig:FandPandNoErrorP}
\end{figure*}

\section{Auxiliary Operations}\label{sec:AuxiliaryOperations}
Now that we have established the approximate state breeding output wave function and homodyne probability distribution, we examine how to improve the protocols' performance. Two auxiliary operations are introduced to increase the overall protocol success probability and fidelity of the output state to the GKP qubit \cite{GKPQubit}. 

The condition on squeezing in Eq. \eqref{eq:SpacingCondition} places a restriction on the rate that optimal GPS states can be generated, hence limiting the overall success probability of the approximate state breeding protocol. The first auxiliary operation lifts this restriction and is done by placing an inline squeezer in the undetected mode of the GPS circuit. With this addition, we can choose the squeezing of the input SV states and inline squeezer such that the probability of detecting $n$ photons is maximized, while the output is still the GPS state with squeezing $r_{\text{opt}}$. Specifically, if the GPS input SV states have squeezing,
\begin{align}
    r_{\text{max}}(n) = \frac{1}{2} \cosh^{-1}\left(1+2 n\right),
    \label{eq:rProbmax}
\end{align}
an inline squeezer in the undetected output mode with 
\begin{align}
    r_c(n) = -\frac{1}{2} \ln \left( \frac{2 \pi (1+2 n)}{n}\right).
\end{align}
will still yield $ \psi^{\text{\tiny{GPS}}}_{r_{\text{\tiny{opt}}}}(q|n)$. Since this inline squeezing is unconditional, we can use Bloch-Messiah decomposition to find an equivalent circuit with all squeezing isolated at the input. This modification improves the  optimal GPS state generation success probability by about two orders of magnitude. The second auxiliary operation is the probabilistic damping operation of \cite{takase2023gottesman}. It is implemented with an ancilla SV state, Gaussian optics, and heralded by on/off detection. We use the operation to improve the fidelity of the breeding output states relative to the GKP qubit. With a successful off detection, this operation, denoted $\mathcal{D}_{r_\text{\tiny{d}}}$, will suppress the breeding output state side peak amplitudes relative to the center peak by a factor $\exp\left[(1- \tanh(r_\text{\tiny{d}})) \pi \right]$, where $r_\text{\tiny{d}}$ is the squeezing of the ancilla SV state.

\section{Results}\label{sec:Results}
We present the analysis of the quality of the states generated from breeding the optimal GPS states in two cases: with and without postselection. When post selection is used, we numerically find
\begin{align}
    r_{\text{\tiny{d}}}^{\text{\tiny{opt}}},~\xi^{\text{\tiny{opt}}} & =\argmax_{r_{\text{\tiny{d}}},\xi ~\in~ \mathbb{R}_{\tiny{\geq 0}}} ~ F_{n}(r_{\text{\tiny{d}}},\xi),\label{eq:Fargmax}\\
    F_{n}(r_{\text{\tiny{d}}},\xi) & = \left| \int dq ~ \mathcal{D}_{r_\text{\tiny{d}}}\left[\phi_{n}\right](q|p_{\text{\tiny{opt}}}) \psi^{\text{\tiny{GKP}}}_{0,\xi}(q)\right|^{2},
\end{align}
for $n \in \{7,\ldots,40\}$, where $p_{\text{\tiny{opt}}}$ is defined in Appendix \ref{app:HomodyneProbability}. $\psi^{\text{\tiny{GKP}}}_{0,\xi}(q)$ is the position wave function of the logical-zero finite-squeezed GKP qubit, with squeezing $\xi$ defined in \cite{GKPQubit}. The result is the amount that each $\phi_{n}(q|p_{\text{\tiny{opt}}})$ should be damped by and the particular GKP qubit, $\psi^{\text{\tiny{GKP}}}_{0,\xi}(q)$, that we should compare them with to get the maximum possible fidelity allowed by our methods. Evidently, it is optimal to apply no damping to the breeding output states with $n \geq 27$. This is because the off-axis peaks of these states are already smaller in amplitude than the off-axis peaks of the GKP qubit with squeezing $\xi = r_{\text{\tiny{opt}}}$. Therefore, we do not apply damping to these states and the optimization in Eq.\eqref{eq:Fargmax} is over $\xi$ only. The probability of generating these states is simply the product of the GPS state generation probability squared, the homodyne measurement success probability, and the damping operation success probability if applicable. 
That is,
\begin{align}
    P^{\text{\tiny{Total}}}_n = \left( P^{\text{\tiny{GPS}}}_{n}(r_{\text{\tiny{max}}}) \right)^2 \cdot P_n^{\text{\tiny{Sum}}} \cdot P_n^{\text{\tiny{Damp}}}(r_{\text{\tiny{d}}}^{\text{\tiny{opt}}}),
\end{align}
where $P_n^{\text{\tiny{Sum}}}$ is the homodyne measurement success probability, defined in Appendix \ref{app:HomodyneProbability}, that takes into account the finite-resolution of the measurement. In Fig. \ref{fig:FandPandNoErrorP}(a) and \ref{fig:FandPandNoErrorP}(b), $F_{n}(r_{\text{\tiny{d}}}^{\text{\tiny{opt}}},~\xi^{\text{\tiny{opt}}})$ and $P^{\text{\tiny{Total}}}_n$ are plotted as a function of $\xi$. Additionally, we calculate the quantity $P_{\text{\tiny{no error}}}$ for our generated states \cite{glancy2006error}. This is the probability that the quantum error-correction scheme of Glancy-Knill succeeds using our generated states. These results and their comparison to the $P_{\text{\tiny{no error}}}$ of finite-squeezed GKP states are shown in Fig. \ref{fig:FandPandNoErrorP}(c). Note that the entire optical circuit that generates these states (the GPS state generation, breeding, and damping operation) can be simplified to an optical circuit with SV inputs operated on with passive Gaussian optics using Bloch-Messiah decomposition.

\begin{figure}
    \centering
    \includegraphics[width=0.9\linewidth]{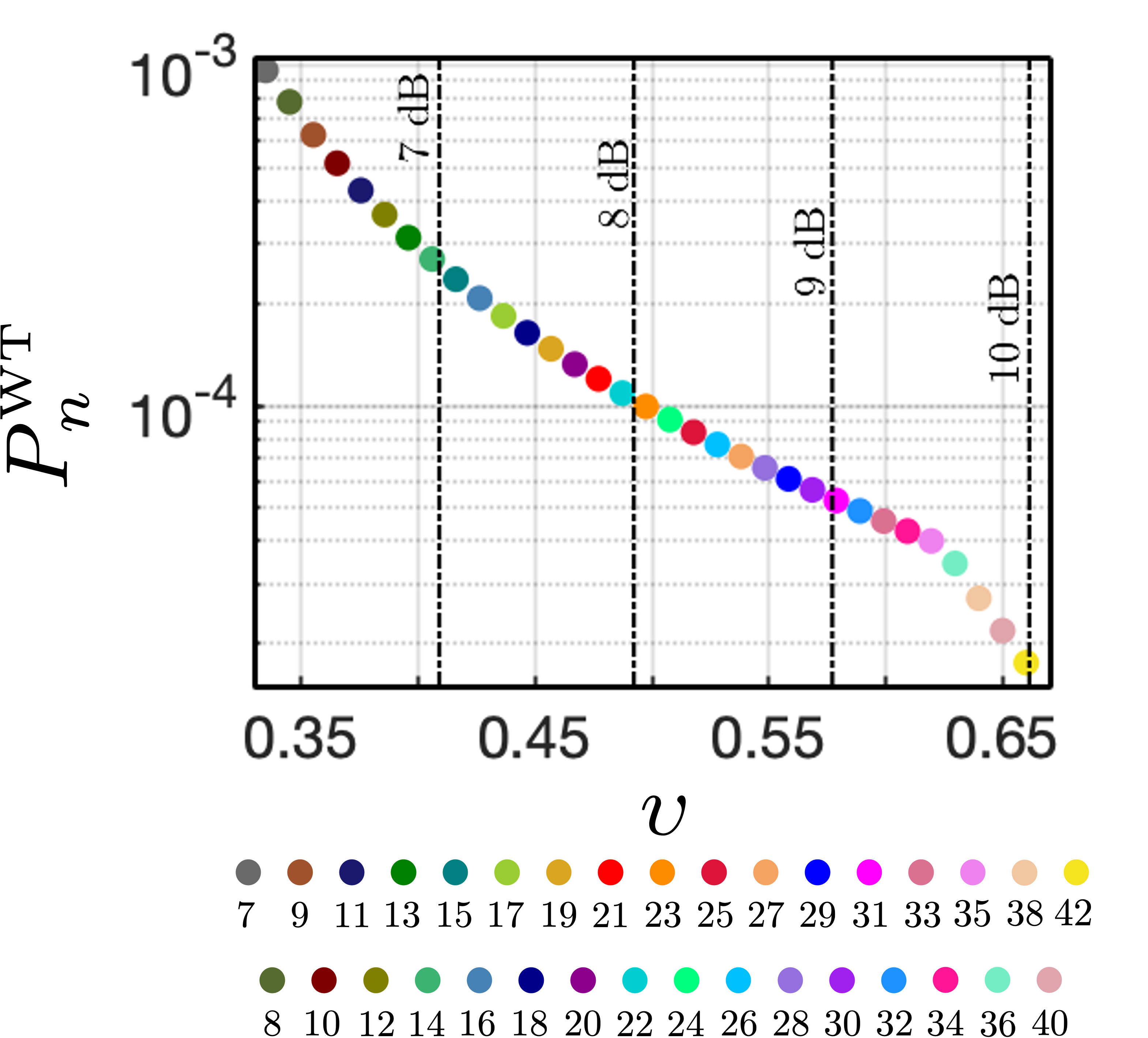}
    \caption{Probability $P^{\text{\tiny{WT}}}_n$ that a state generated by breeding without postselection and appropriate corrective displacement, $\phi_{n,\delta}(q_1|p)$, will have a $P_{\text{\tiny{no error}}} \geq \upsilon$. The vertical dash-dotted lines mark the $P_{\text{\tiny{no error}}}$ of a logical-zero finite-squeezed GKP qubit, with $\xi$ equal to the squeezing value each line is labeled with. The color of each point denotes the number of photons, $n$, associated with the GPS states used to create the states $\phi_{n,\delta}(q_1|p)$.}
    \label{fig:PvsNoErrorP}
\end{figure}

When breeding without postselection, the damping operation would require in-line squeezing since the Bloch-Messiah decomposition of the circuit depends on the $r_{\text{\tiny{d}}}^{\text{\tiny{opt}}}$, which, in turn depends on, the homodyne measurement outcome. To maintain experimental simplicity, we forego the damping operation in this case. Since the $P_{\text{\tiny{no error}}}$ of a state, rather than fidelity to $\psi^{\text{\tiny{GKP}}}_{0,\xi}(q)$, directly determines the performance of said state in the Glancy-Knill error-correction scheme, we report the total probability that a state $\phi_{n,\delta}(q_1|p)$ will have a $P_{\text{\tiny{no error}}} \geq \upsilon$ as
\begin{align}
    P^{\text{\tiny{WT}}}_n = \left( P^{\text{\tiny{GPS}}}_{n}(r_{\text{\tiny{max}}}) \right)^2 \cdot  P^{\upsilon}.
\end{align}
For a given $\phi_{n,\delta}(q_1|p)$, $P^{\upsilon}$ is found by determining the regions of $p \in \mathbb{R}$ over which $\phi_{n,\delta}(q_1|p)$ has $P_{\text{\tiny{no error}}} \geq \upsilon$ and then integrating $P_n^{\text{\tiny{Hom}}}(p)$ over these regions. These results are plotted in Fig. \ref{fig:PvsNoErrorP} and show that states with a $P_{\text{\tiny{no error}}}$ equivalent to that of a $\xi_{\text{\tiny{dB}}} = 10 ~\text{dB} $ logical-zero finite-squeezed GKP qubit can be generated with success probability $\geq 10^{-5}$ using approximate breeding without postselection.

\section{Discussion}\label{sec:Discussion}
Our work is a comprehensive analysis of the state breeding protocol for GKP qubit preparation that takes into account imperfect input states and their generation probability. When trying to prepare GKP qubits beyond $10~\text{dB}$ of squeezing, our method will have diminishing returns due to the fact that GKP qubits with that level of squeezing have more than three Gaussian peaks in their wave function. Nonetheless, $10~\text{dB}$ is often considered the threshold for fault-tolerant quantum computation \cite{fukui2018high} and our scheme performs exceptionally well in preparing approximate GKP qubits with squeezing equal to this mark. Of course, more peaks can be added to our states with additional rounds of breeding, but this would require generating more GPS states diluting the success probability of the overall scheme, as pointed out in \cite{takase2023gottesman}. This problem was recently addressed by introducing adaptive inline squeezing at the GPS circuit output \cite{Takase_2024}. By conditioning on the PNR measurement, any photon number outcome above a certain threshold could be accepted and the peak spacing of the heralded state corrected by the inline squeezer before breeding. This modification to the GPS scheme allowed for multiple breeding rounds to be carried out without dilution of the success probability. While the target states in \cite{Takase_2024} are GKP states with $\sqrt{2 \pi}$ peak spacing, called qunaught states \cite{Walshe_2020,duivenvoorden2017single}, our work here, where we target a GKP qubit state with $2 \sqrt{\pi}$ peak spacing, may be seen as the limiting case of \cite{Takase_2024} without adaptive inline squeezing. 

State breeding is a well known idea; therefore their is already a body of experimental work on the subject. For example, the use of state breeding to enlarge cat states has already been experimentally demonstrated \cite{sychev2017enlargement}. A notable recent experiment was able to breed a pair of small SCSSs to create a state with three Gaussian peaks \cite{konno2024logical}.
Additionally, a nonlinear feed-forward operation has been demonstrated in \cite{sakaguchi2023nonlinear} that could be used to perform the homodyne measurement-based corrective displacement utilized in this work. State-of-the-art PNR detectors based on arrays of superconducting nanowire single-photon detectors with resolution up to 24 photons can operate in the hundreds-of-MHz regime \cite{Mattioli2016}. With improvements to the photon number resolution of these detectors and just 1000 of our sources in parallel, one could generate $ 10$ dB squeezed GKP qubits at a rate of MHz.

While GPS can generate approximate SCSSs at a high rate, it remains to be seen if it is the most efficient method. Furthermore, when using breeding to approximate GKP qubits, there is no consensus that SCSSs are the optimal inputs. In \cite{zheng2023gaussian}, the authors studied breeding binomial states to generate qunaught states. The binomial state generation scheme of \cite{nehra2022all} could be used to do a rigorous performance analysis of binomial state breeding.

\section{Conclusion}\label{sec:Conclusion}
We have shown that state breeding with and without postselection can generate high-quality approximations of the finite-squeezed GKP qubits with squeezing up $10~ \text{dB}$ with a success probability nearly two orders of magnitude greater than the best-known strictly PNR measurement-based methods. This was done by limiting our protocol to two PNR measurements supplemented by a single homodyne measurement. In addition to having fidelities $\geq 0.98$ to the finite-squeezed GKP qubits up to $\xi_{\text{\tiny{dB}}} = 10 ~\text{dB}$, the states generated with postselection perform nearly as well as their equivalent finite-squeezed GKP qubit in the Glancy-Knill error-correction scheme. After obtaining the functional form of the GPS breeding output state displaced in its momentum quadrature and the homodyne measurement-dependent corrective displacement amplitude, we analyzed the GPS states in the context of breeding without post selection. Specifically, we calculated the probability that the state produced by GPS state breeding without postselection results in a state with a Glancy-Knill no-error probability above a given threshold. We found that the threshold associated with a logical-zero finite-squeezed GKP qubit with $\xi = 10 ~\text{dB}$ is attainable with a probability $\geq 10^{-5}$. We estimate that with this marked improvement in success probability and near term PNR detection hardware, MHz generation of 10 dB squeezed GKP qubits is feasible. 

\section{Data Availability}
The data and code that support the findings of this article are openly available \cite{Ajpizzimenti2024-uz}.

\section{Acknowledgments}\label{sec:Acknowledgments}
We would like to thank Joseph Lukens and Christos Gagatsos for helpful discussions.

\appendix

\section{General State Breeding}\label{sec:GeneralStateBreeding}
\begin{figure}
    \centering
    \includegraphics[width=0.75\linewidth]{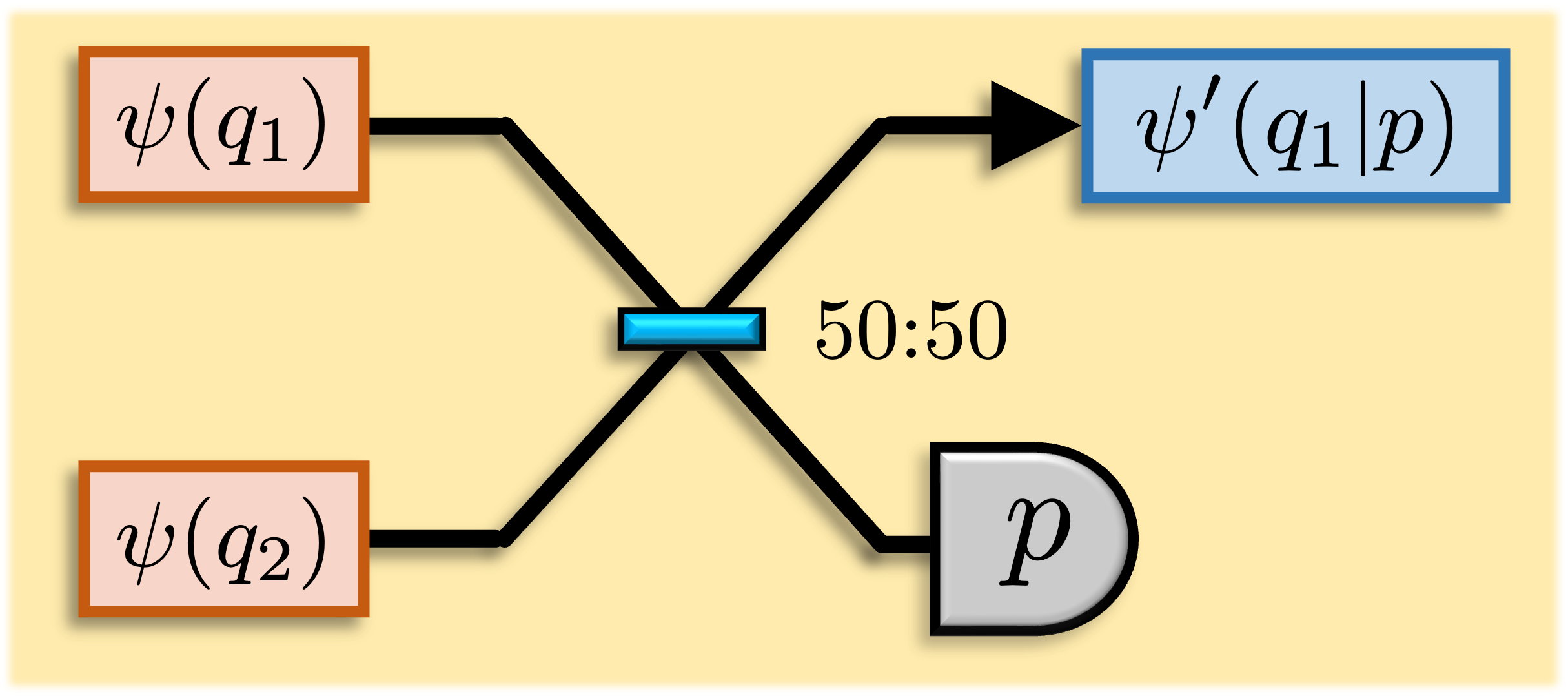}
    \caption{General breeding with two copies of the state $\psi$. The two states are combined on a $50$:$50$ beam splitter and a homodyne measurement in mode $2$ heralds the state $\psi'(q_1|p)$ in mode $1$.}
    \label{fig:GeneralBreedingAndGPS}
\end{figure}

We work in units where $\hbar =1$ such that the position and momentum operators obey the commutation relation $[\hat{q},\hat{p}] = i$. We define the displacement operator to be $\hat{D}(\alpha) = e^{\alpha \hat{a}^{\dagger}- \alpha^* \hat{a}}$ $(\alpha \in \mathbb{C})$ and squeezing operator as $\hat{S}(\xi) = e^{\frac{\xi}{2}(\hat{a}^{\dagger 2}-\hat{a}^{2})}$ $(\xi \in \mathbb{R})$. The strategy of state breeding is a well-known idea \cite{vasconcelos2010, weigand2018}, but here we give a brief overview of the general breeding process, shown in Fig. \ref{fig:GeneralBreedingAndGPS}, using position wave functions. We start with two copies of an arbitrary state $\ket{\psi}$ in a product state,
\begin{align}
    \ket{\psi}_1 \otimes \ket{\psi}_2 = \int \int dq_1 dq_2~ \psi(q_1) \psi(q_2) \ket{q_1}_1 \ket{q_2}_2.
    \label{eq:BreedingStep1}
\end{align}
The balanced beam splitter transforms the coordinates as
\begin{align}
    q_1  \rightarrow \frac{q_1+q_2}{\sqrt{2}},~~  & q_2  \rightarrow \frac{q_2-q_1}{\sqrt{2}} 
\end{align}
resulting in an entangled state
\begin{align}
    \ket{\psi'}_{1,2} = \int \int dq_1 dq_2 ~\psi'(q_1,q_2) \ket{q_1} \ket{q_2},
\end{align}
where we have omitted the mode indices outside the position eigenstate kets as the mode it corresponds to is implicit in the variable $q_{1,2}$. When a homodyne measurement is performed on the momentum quadrature of mode $2$ with outcome $p$, the unnormalized state heralded in mode $1$ is found by projecting $|\psi' \rangle_{1,2}$ on the momentum eigenstate of mode $2$, $\ket{p}$,
\begin{align}
    \langle p|\psi' \rangle_{1,2}  & = \int \int dq_1 dq_2  ~ \psi'(q_1,q_2) ~\langle p  \ket{q_2}~\ket{q_1}.
\end{align}
Resolving the inner product as $~\langle p  \ket{q_2} = e^{ip q_2}/\sqrt{2 \pi}$, the unnormalized wave function is
\begin{align}
    \Psi'(q_1|p)= \frac{1}{\sqrt{2 \pi}}\int dq_2~\psi'(q_1,q_2) e^{ip q_2}.
\end{align}
The probability associated with outcome $p$ is simply the normalization constant of $\Psi'(q_1|p)$ squared
\begin{align}
    P(p) & = \int dq_1 \left| \Psi'(q_1|p) \right|^2 ,\\
    \nonumber \\
    \psi'(q_1|p) & = \frac{1}{\sqrt{P(p_2)}} \Psi'(q_1|p).
\end{align}
When modeling a postselected homodyne measurement one must take into account the finite-resolution of the measurement. The positive operator-valued measure (POVM) for a momentum homodyne measurement, post-selected on an outcome $\tilde{p}$, with a finite-resolution width of $2 \epsilon$ is \cite{Zheng_2021,Paris_2003,Douce_2017} 
\begin{align}
    \hat{\Pi}_{\tilde{p}}(\epsilon) & = \int_{\tilde{p}-\epsilon}^{\tilde{p}+\epsilon} dp \ket{p}\bra{p}.
\end{align}
Performing this measurement on the second mode of a two-mode entangled state $\ket{\psi'}_{1,2}$ gives
\begin{align}
    & \Tr_2\left[ \left(\hat{\mathbbm{1}}_1  \otimes \hat{\Pi}_{\tilde{p},2}(\epsilon)  \right)\ket{\psi}_{1,2} \bra{\psi}_{1,2}  \left(\hat{\mathbbm{1}}_1  \otimes \hat{\Pi}_{\tilde{p},2}(\epsilon)  \right) \right] = \nonumber 
    \\
    & \frac{1}{2 \pi} \int_{\tilde{p}-\epsilon}^{\tilde{p}+\epsilon} dp \int \int dq_1 dq_2 \int \int dq_1' dq_2'~ \psi'\left(q_1,q_2\right) \psi'^{*}\left(q_1',q_2'\right) \nonumber 
    \\
    & \times e^{i q_2 p} e^{-i q_2' p} \ket{q_1}_1 \bra{q_1'}_1.
\end{align}
The unnormalized conditional output state function is then 
\begin{align}
\Gamma'\left(q_1,q_1',\epsilon|\tilde{p}\right) & =  \frac{1}{2 \pi} \int_{\tilde{p}-\epsilon}^{\tilde{p}+\epsilon} dp \int \int dq_2 dq_2'~ \psi'\left(q_1,q_2\right) \nonumber 
\\
 & \times \psi'^{*}\left(q_1',q_2'\right) e^{i q_2 p} e^{-i q_2' p} \nonumber 
 \\
& = \frac{1}{2 \pi} \int_{\tilde{p}-\epsilon}^{\tilde{p}+\epsilon} dp ~ \Psi'\left(q_1|p\right) \Psi'^{*}\left(q_1'|p\right).
\end{align}
The probability of obtaining the outcome $\tilde{p}$ is the normalization constant of $\Gamma'\left(q_1,q_1'|\tilde{p}\right)$,
\begin{align}
    P\left(\epsilon,\tilde{p}\right) & = \Tr_1\left[\hat{\Gamma}'\right] \nonumber 
    \\
    & = \frac{1}{2 \pi} \int_{\tilde{p}-\epsilon}^{\tilde{p}+\epsilon} dp \int dq_1 ~ \left| \Psi'\left(q_1|p\right) \right|^2. 
    \\
    \nonumber 
    \\
    \rho\left(q_1,q_1',\epsilon|\tilde{p}\right) & = \frac{1}{P\left(\epsilon,\tilde{p}\right)} \Gamma'\left(q_1,q_1',\epsilon|\tilde{p}\right). \label{eq:rhocon}
\end{align}
We will return to this finite-resolution method when discussing the overall success probability of approximate state breeding with postselection.

\section{\texorpdfstring{$\phi_{n}(q_1|p)$}{TEXT} Derivation}
\label{app:phi_n}
The GPS state wave function with optimal input squeezing,
\begin{align}
    r_{\text{opt}}(n) = \frac{1}{2} \sech^{-1}\left(\frac{2 \pi}{n}\right),
\end{align}
is given by \cite{takase2021generation}  
\begin{align}
     \psi^{\text{\tiny{GPS}}}_{r_{\text{\tiny{opt}}}}(q|n) = \frac{1}{\sqrt{\Gamma\left(\frac{1}{2}+n\right)}} \left(\frac{n}{2 \pi}\right)^{\frac{2n+1}{4}} \left(-q \right)^n e^{-\frac{1}{2}\frac{n}{2 \pi} q^2}.
    \label{eq:GPSStateOptSqueezing}
\end{align}
Taking two copies of this state and performing the beam-splitter transformation leads to 
\begin{align}
    \psi'_{n}(q_1,q_2) & = \frac{(-1)^n}{\Gamma\left(\frac{1}{2}+n\right)} \frac{n^{\frac{2n+1}{2}}}{2^{2n+1/2}\pi^{n+1/2}}e^{-\frac{1}{2}\frac{n}{2 \pi} \left(q_1^2 +q_2^2\right)} \nonumber \\
    & \times \left(q_1-q_2\right)^n \left(q_1+q_2\right)^n.
    \label{eq:GPSBreedingPreDetectionState}
\end{align}
Before continuing in our derivation, we rewrite the last two factors in Eq. \eqref{eq:GPSBreedingPreDetectionState} as
\begin{align}
    \left(q_1-q_2\right)^n \left(q_1+q_2\right)^n & = q_1^{2n} \left(1-\frac{q_2}{q_1}\right)^n \left(1+\frac{q_2}{q_1}\right)^n \nonumber
    \\
    & =
    q_1^{2n} \left[1 -\left(\frac{q_2}{q_1}\right)^2 \right]^n \nonumber
    \\
    & =
    q_1^{2n} \sum_{k = 0}^n (-1)^k \binom{n}{k} \left(\frac{q_2}{q_1}\right)^2 \nonumber
    \\
    & = 
    \sum_{k = 0}^n (-1)^k \binom{n}{k} q_1^{2(n-k)} q_2^{2k},
\end{align}
where we have used the binomial expansion going from the second to the third line. The unnormalized state after homodyne detection with infinite-resolution is then given by
\begin{align}
       \Psi'_{n}(q_1|p) & = \frac{1}{\Gamma\left(\frac{1}{2}+n\right)} \left(\frac{-n}{4 \pi}\right)^n \frac{\sqrt{n}}{2 \pi}
       e^{-\frac{1}{2}\frac{n}{2 \pi} q_1^2 } \nonumber 
       \\
       & \times \sum_{k = 0}^n (-1)^k \binom{n}{k} q_1^{2(n-k)} \int dq_2 ~ q_2^{2k} e^{-\frac{1}{2}\frac{n}{2 \pi} q_2^2}e^{i p q_2}.
       \label{eq:GPSBreedingPostHomodyne}
\end{align}
An integral of this form in Eq.\eqref{eq:GPSBreedingPostHomodyne} will show up multiple times in our analysis; therefore we solve it generally here,
\begin{align}
    \Theta_k(y, \gamma) & =  \int_{-\infty}^{\infty} dx ~ x^{2k} e^{-\frac{\gamma}{2} x^2}e^{i y x}.
\end{align}
First, we rewrite the expression by completing the square in the argument of the exponential,
\begin{align}
    \Theta_k(y, \gamma) & =   \int_{-\infty}^{\infty} dx ~ x^{2k} e^{-\frac{\gamma}{2} x^2}e^{i y x} \nonumber 
    \\
    & = e^{-\frac{y^2}{2 \gamma}} \int_{-\infty}^{\infty} dx ~ x^{2k} e^{-\frac{\gamma}{2} \left( x - i\frac{y}{\gamma}\right)^2} .
\end{align}
Making the change of variables $u = x -  i\frac{y}{\gamma}$,
\begin{align}
    \Theta_k(y, \gamma) &  = e^{-\frac{y^2}{2 \gamma}}  \int_{-\infty}^{\infty} dx ~ \left( u + i\frac{y}{\gamma} \right)^{2k} e^{-\frac{\gamma}{2}u^2} .
\end{align}
Using the binomial expansion, this is rewritten as
\begin{align}
    \Theta_k(y, \gamma) &  = e^{-\frac{y^2}{2 \gamma}} \sum_{l = 0}^{2 k} \binom{2 k}{l} \left(  i\frac{y}{\gamma}\right)^{2 k - l}  \int_{-\infty}^{\infty} dx ~u^l e^{-\frac{\gamma}{2}u^2} .
    \label{eq:Thetaintermediate}
\end{align}
The integral in Eq.\eqref{eq:Thetaintermediate} is the general moments of a zero-mean Gaussian,
\begin{align}
     \int_{-\infty}^{\infty} dx ~u^l e^{-\frac{\gamma}{2}u^2} & =  \frac{\left[ 1 + \left(-1\right)^l \right]}{\sqrt{2 \gamma}} \left(\frac{2}{\gamma}\right)^{\frac{l}{2}}  \Gamma\left( \frac{1 + l}{2}\right).
\end{align}
Simplifying the expression in Eq.\eqref{eq:Thetaintermediate},
\begin{align}
    \Theta_k(y,\gamma) &  = \int_{-\infty}^{\infty} dx ~ x^{2k} e^{-\frac{\gamma}{2} x^2}e^{i y x} \nonumber 
    \\
    & = e^{-\frac{y^2}{2 \gamma}} \sum_{l = 0}^{2 k} f^k_l(\gamma)~ y^{2(k-l)},
    \label{eq:AppThetaFinal}
\end{align}
where
\begin{align}
    f^k_l(\gamma) & =  \left( -\frac{1}{\gamma^2} \right)^{k-l} \binom{2 k}{2 l} \left( \frac{2}{\gamma} \right)^{l+\frac{1}{2}} \Gamma\left( l + \frac{1}{2} \right) .
\end{align}
Substituting  $\Theta_k(p,n/2 \pi)$ into Eq.\eqref{eq:GPSBreedingPostHomodyne} and simplifying we find 
\begin{align}
    \Psi'_{n}(q_1|p) & = \frac{1}{  \Gamma\left(\frac{1}{2}+n\right)} \left(\frac{-n}{4 \pi}\right)^n \frac{\sqrt{n}}{2 \pi} e^{-\frac{1}{2}\frac{n}{2 \pi} q_1^2 } \nonumber \\
    & \times \sum_{k = 0}^n  ~\Theta_k\left(p,\frac{n}{2 \pi}\right) ~ (-1)^k \binom{n}{k}q_1^{2(n-k)} 
\end{align}
The homodyne probability distribution is 
\begin{align}
    P_n^{\text{\tiny{Hom}}}(p) & = \left[\frac{1}{  \Gamma\left(\frac{1}{2}+n\right)} \left(\frac{-n}{4 \pi}\right)^n \frac{\sqrt{n}}{2 \pi} \right]^2 \times \mathcal{N}^2
\end{align}
and the final normalized state is
\begin{align}
    \phi_{n}(q_1|p) &  = \frac{e^{-\frac{1}{2}\frac{n}{2 \pi} q_1^2 }}{\mathcal{N}} \sum_{k = 0}^n ~\Theta_k\left(p,\frac{n}{2 \pi}\right) ~(-1)^k \binom{n}{k}q_1^{2(n-k)},
\end{align}
where
\begin{widetext}
    \begin{align}
        \mathcal{N} & = \sqrt{\sum_{k,k'=0}^n ~\Theta_k\left(p,\frac{n}{2 \pi}\right) ~\Theta_{k'}\left(p,\frac{n}{2 \pi}\right)  (-1)^{k+k'} \binom{n}{k} \binom{n}{k'} \left(\frac{2 \pi}{n}\right)^{2n-k-k'+\frac{1}{2}} \Gamma\left( 2n-k-k'+\frac{1}{2} \right)}.
    \end{align}
\end{widetext}

\section{\texorpdfstring{$\phi_{n}(p_1|p)$}{TEXT}, \texorpdfstring{$\delta$}{TEXT}, and \texorpdfstring{$\phi_{n,\beta}(q_1|p)$}{TEXT} Derivation}
\label{app:CorrectiveDispacmentAndOutputState}
According to Ref. \cite{weigand2018}, one can determine the corrective displacement needed by first calculating the mean phase $\theta$ of the breeding output state $ \ket{\psi'}$,
\begin{align}
    \theta = \text{arg}~ \bra{\psi'} \hat{D}(\sqrt{2 \pi}) \ket{\psi'}.
\end{align}
For the ideal GKP qubit, $\theta = 0$. Displacing the breeding output state by an amount $\delta / \sqrt{2}$, such that $\hat{D}(\sqrt{2 \pi}) \hat{D}(i \delta / \sqrt{2}) = e^{- i \theta}  \hat{D}(i \delta / \sqrt{2}) \hat{D}(\sqrt{2 \pi})$, yields a state with $\theta \approx 0$. Working out the commutation relation between $\hat{D}(\sqrt{2 \pi})$ and $\hat{D}(i \delta / \sqrt{2})$, we find
\begin{align}
    \delta = \frac{\theta}{2\sqrt{\pi}}.
    \label{eq:delta}
\end{align}
To calculate the mean phase $\theta$ of $\phi_{n}(q_1|p)$, we use the momentum wave function so we can take advantage of the fact that $\hat{D}(\sqrt{2 \pi}) \ket{p_1} = e^{i 2 \sqrt{\pi} p_1} \ket{p_1}$. The momentum-space wave function of the GPS breeding output state is simply the Fourier transform of $\phi_{n}(q_1|p)$, 
\begin{align}
    \phi_{n}(p_1|p) & = \frac{1}{\sqrt{2 \pi}} \int dq_1 ~ \phi_{n}(q_1|p) e^{-i q_1 p_1} \nonumber 
    \\
    & = \frac{1}{ \sqrt{2 \pi}  \mathcal{N}}  \sum_{k = 0}^n ~\Theta_k\left(p,\frac{n}{2 \pi}\right) (-1)^{k} \binom{n}{k} \nonumber 
    \\
   & \times \int dq_1 ~ q_1^{2(n-k)}  e^{-\frac{1}{2}\frac{n}{2 \pi} q_1^2 } e^{-i q_1 p_1}.
\end{align}
This integral is of the form $\Theta_{n-k}(p_1, n/ 2 \pi)$; therefore we can substitute in Eq. \eqref{eq:AppThetaFinal} and simplify to find 
\begin{align}
    \phi_{n}(p_1|p) & = \frac{e^{-\frac{1}{2} \frac{2 \pi }{n} p_1^2}}{ \sqrt{2 \pi} \mathcal{N} }  \sum_{k = 0}^n \sum_{j=0}^{2(n-k)} ~\Theta_k\left(p,\frac{n}{2 \pi}\right) (-1)^{k} \binom{n}{k} \nonumber 
    \\
    & \times f^{n-k}_j\left(\frac{n}{2 \pi}\right)~ p_1^{2(n-k-j)}.
    \label{eq:TildePhiFT}
\end{align}
The mean phase of the state $\phi_{n}(p_1|p)$ is 

\begin{align}
    \theta & = \text{arg} \int dp_1 ~ \left| \phi_{n}(p_1|p) \right|^2 e^{i 2 \sqrt{\pi} p_1} \nonumber
    \\
     & =\text{arg } \frac{1}{2 \pi \mathcal{N}} \sum_{k,k'=0}^{n} \sum_{j = 0}^{2(n-k)} \sum_{j' = 0}^{2(n-k')} (-1)^{k+k'} \nonumber 
     \\
     & \times\Theta_{k}\left(p,\frac{n}{2 \pi}\right) \Theta_{k'}\left(p,\frac{n}{2 \pi}\right) 
     \binom{n}{k} \binom{n}{k'} \nonumber 
     \\
     & \times f_{j}^{n-k}\left(\frac{n}{2 \pi}\right) ~f_{j'}^{n-k'}\left(\frac{n}{2 \pi}\right) \nonumber
     \\
     & \times \int dp_1 ~p_1^{2(2n-\Delta k - \Delta j)} e^{- \frac{2 \pi}{n} p_1^2} e^{i 2 \sqrt{\pi} p_1},
\end{align}

where $\Delta k = k - k'$ and $\Delta j = j - j'$. The integral is of the form $\Theta_{2n-\Delta k - \Delta j}(2 \sqrt{\pi}, 4 \pi /n)$; therefore,
\begin{align}
    \theta & = \text{arg } g_n(p) \nonumber\\
    g_n(p) & =  \frac{1}{2 \pi \mathcal{N}} \sum_{k,k'=0}^{n} \sum_{j = 0}^{2(n-k)} \sum_{j' = 0}^{2(n-k')}  (-1)^{k+k'} \binom{n}{k} \binom{n}{k'} \nonumber \\
    & \times \Theta_{k}\left(p,\frac{n}{2 \pi}\right) \Theta_{k'}\left(p,\frac{n}{2 \pi}\right) \Theta_{2n-\Delta k - \Delta j}\left(2 \sqrt{\pi}, \frac{4 \pi}{n}\right) \nonumber 
    \\ 
    & \times f_{j}^{n-k}\left(\frac{n}{2 \pi}\right) ~f_{j'}^{n-k'}\left(\frac{n}{2 \pi}\right)
    \label{eq:thetaAnalytic}
\end{align}
The function $g_{n}(p)$ is a continuous function of $p$ that can take on positive or negative values for different regions of $p$ depending on $n$. Therefore,
\begin{align}
    \theta(p) & =  \begin{cases}
         0, &  g_{n}(p) > 0 \\
         \pi, &  g_{n}(p) < 0 \\
         \text{undefined}, & g_{n}(p) = 0 
    \end{cases},
\end{align}
and the choice of corrective displacement amplitude amounts to a binary decision 
\begin{align}
    \delta & =  \begin{cases}
         0, &  g_{n}(p) \geq 0 \\
         \frac{\sqrt{\pi}}{2}, &  g_{n}(p) < 0 \\
    \end{cases}.
    \label{eq:deltaBinary}
\end{align}
Displacing the state $\phi_{n}(p_1|p)$ by $\beta/\sqrt{2}$ in the momentum quadrature yields the state $\phi_{n}(p_1+\beta|p)$. By taking the inverse Fourier transform of this state to get back the position wave function, 
\begin{align}
    \phi_{n,\beta}(q_1|p) & = \frac{1}{\sqrt{2 \pi}} \int dp_1 ~ \phi_{n}(p_1 + \beta|p) e^{i q_1 p_1} \nonumber 
    \\
    & = \frac{1}{2 \pi \mathcal{N}}  \sum_{k = 0}^n \sum_{j=0}^{2(n-k)}  ~\Theta_k\left(p,\frac{n}{2 \pi}\right) \nonumber 
    \\
    & \times (-1)^{k} \binom{n}{k}  ~f^{n-k}_j\left(\frac{n}{2 \pi}\right) \nonumber 
    \\
    & \times \int dp_1 ~\left(p_1+\beta\right)^{2(n-k-j)} e^{-\frac{1}{2} \frac{2 \pi }{n} \left(p_1+\beta\right)^2}  e^{i q_1 p_1}.
\end{align}
By making the change of variables  $u = p_1 + \beta$,
\begin{align}
    \phi_{n,\beta}(q_1|p) & =  \frac{e^{ -i q_1 \beta}}{2 \pi \mathcal{N}} \sum_{k = 0}^n \sum_{j=0}^{2(n-k)}  ~\Theta_k\left(p,\frac{n}{2 \pi}\right) \nonumber 
    \\
    & \times (-1)^{k} \binom{n}{k}  ~f^{n-k}_j\left(\frac{n}{2 \pi}\right) \nonumber 
    \\
    & \times \int du ~u^{2(n-k-j)} e^{-\frac{1}{2} \frac{2 \pi }{n} u^2}  e^{i q_1 u} .
\end{align}
The integral is of the form $\Theta_{n-k-j}(p_1, 2 \pi /n)$; therefore,
\begin{align}
    \phi_{n,\beta}(q_1|p) & =  \frac{e^{-\frac{1}{2}\frac{n}{2 \pi} q_1^2  -i q_1 \beta}}{2 \pi \mathcal{N}} \sum_{k = 0}^n \sum_{j=0}^{2(n-k)} \sum_{m=0}^{2(n-k-j)} (-1)^{k} \binom{n}{k} \nonumber 
    \\ 
    & \times f^{n-k}_j\left(\frac{n}{2 \pi}\right) ~ f^{n-k-j}_m\left(\frac{2 \pi}{n}\right) \nonumber 
    \\
    & \times \Theta_k\left(p,\frac{n}{2 \pi}\right) q_1^{2(n-k-j-m)}.
\end{align}

\section{Exact State Breeding}\label{app:ExactStateBreeding}
Before we can compare our approximate state breeding to exact breeding, we must generalize the results of Refs. \cite{vasconcelos2010,weigand2018} to include odd squeezed coherent state superposition (SCSS) breeding.  The SCSS is defined as $\ket{\psi_{\kappa,\alpha,r}^{\text{\tiny{SCSS}}}} \propto  \hat{S}(r)[\hat{D}(\alpha)+(-1)^{\kappa} \hat{D}(-\alpha) ]\ket{0}$ with $ (\kappa \in \mathbb{Z}, \alpha \in \mathbb{R})$. The even(odd) SCSS corresponds to $\kappa ~ \text{mod}~2 = 0(\neq0)$. By breeding even or odd SCSSs with $\alpha_{\text{\tiny{opt}}} = \sqrt{\pi}e^{r}$, one obtains
\begin{align}
    \tilde{\phi}_{\kappa,r}(q_1|p) & \propto \Big[e^{-\frac{e^{2r}}{2}\left(q_1+2\sqrt{\pi}\right)^2}  + e^{-\frac{e^{2r}}{2}\left(q_1+2\sqrt{\pi}\right)^2} \nonumber 
    \\
    & +  2 \cos\left(2 \sqrt{\pi}p + \kappa \pi \right) e^{-\frac{e^{2r}}{2}q_1^2}\Big].\label{eq:ExactBreedingWaveFuinction} 
\end{align}
The ideal output,
\begin{align}
    \tilde{\phi}_{r}(q_1) & =  \frac{e^{-\frac{e^{2r}}{2}\left(q_1+2\sqrt{\pi}\right)^2} + 2 e^{-\frac{e^{2r}}{2}q_1^2} + e^{-\frac{e^{2r}}{2}\left(q_1-2\sqrt{\pi}\right)^2} }{\pi^{\frac{1}{4}}\sqrt{2e^{-r}\left(3+e^{-4e^{2r}\pi}+4e^{-e^{2r}\pi}\right)}}\label{eq:IdealBreedingState},
\end{align}
is heralded with postselection on 
\begin{align}
\tilde{p}_l = \begin{cases} 
       l \sqrt{\pi}, & \kappa ~ \text{is even} \\
       (2l+1)\frac{\sqrt{\pi}}{2}, & \kappa ~ \text{is odd} 
\end{cases},
\label{eq:SSCBreedingConditions}
\end{align}
with $l \in \mathbb{Z}$. The homodyne measurement probability distribution is
\begin{widetext}
\begin{align}
    \tilde{P}_{\kappa,r}^{\text{\tiny{Hom}}}(p) & = \frac{e^{-\frac{e^{-2 r}}{2}(\sqrt{2}p)^2} \Big[1 +e^{4e^{2 r}\pi}\left(2 \cos^{2}\left(2\sqrt{\pi} p \right) +1 \right)
     + 4 e^{3e^{2 r} \pi} \cos\left( 2\sqrt{\pi} p + \kappa \pi\right) \Big]}{2 e^{r}\sqrt{\pi}\left[ (-1)^{\kappa} + e^{2 e^{2 r} \pi} \right]^{2}} \label{eq:ExactBreedingHomodyneDistribution}
\end{align}
\end{widetext}
Note that the outcomes $\tilde{p}_l$ occur at peaks in $\tilde{P}_{\kappa,r}^{\text{\tiny{Hom}}}(p)$.


The choice of corrective displacement amplitude amounts to a similar binary choice as in Eq. \eqref{eq:deltaBinary},
\begin{align}
    \tilde{\delta} & =  \begin{cases}
         0, &  \tilde{g}_{\kappa,r}(p) \geq 0 \\
         \frac{\sqrt{\pi}}{2}, &  \tilde{g}_{\kappa,r}(p) < 0 \\
    \end{cases},
\end{align}
where 
\begin{widetext}
\begin{align}
    \tilde{g}_{\kappa,r}(p) & = \frac{e^{-5 e^{2 r} \pi}+e^{3 e^{2 r} \pi} \left( 5 +2 \cos\left(4 \sqrt{\pi} p\right)  \right) +  4 \left( 1 + e^{4 e^{2 r} \pi} \right) \cos\left( 2 \sqrt{\pi} p + \kappa \pi \right)}{1 + e^{5 e^{2 r} \pi} \left( 2 + \cos \left( 4 \sqrt{\pi} p \right) \right) + 4 e^{3 e^{2r} \pi} \cos \left( 2 \sqrt{\pi} p + \kappa \pi \right)}.
\end{align}
\end{widetext}
A displacement of $\hat{D}(i \beta / \sqrt{2})$, $\beta \in \mathbb{R}$, on the state $\tilde{\phi}_{\kappa,r}(q_1|p)$, manifests as an additional phase term. That is, the position wave function of the exact breeding output state after a displacement in momentum is given by 
\begin{align}
    \tilde{\phi}_{\kappa,r,\beta}(q_1|p) & = e^{-i \beta q_1} \tilde{\phi}_{\kappa,r}(q_1|p).
    \label{eq:DisplacedExactBreedingwave function}
\end{align}

\begin{figure}
    \centering
    \includegraphics[width=0.75\linewidth]{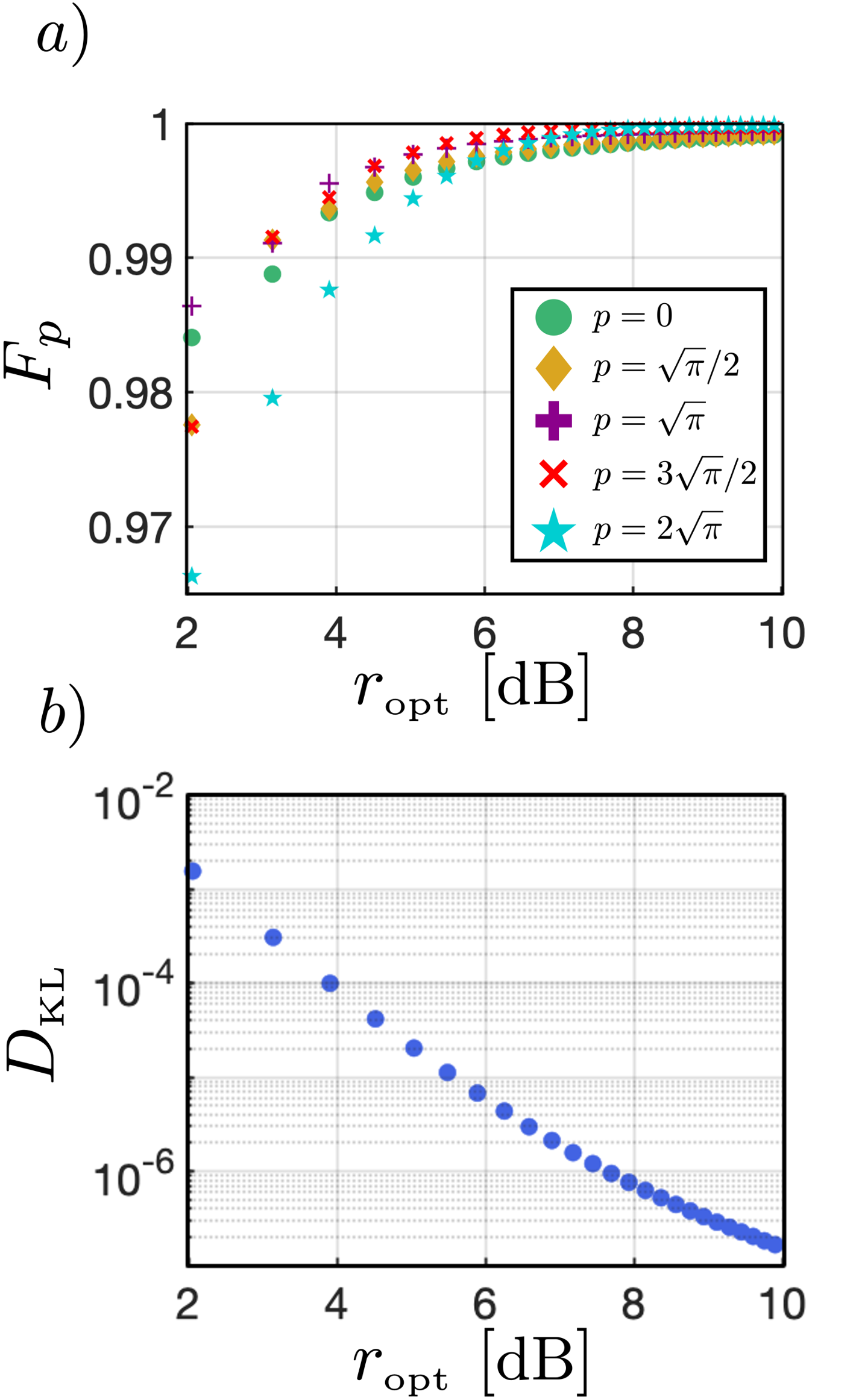}
    \caption{(a) Fidelity between the states that result from exact and approximate breeding when $p \in \{0,~\sqrt{\pi}/2,~ \sqrt{\pi},~3\sqrt{\pi}/2,~ 2\sqrt{\pi}\}$ and the corresponding optimal corrective displacements are applied. (b) Kullback-Leibler divergence between the homodyne probability distributions that correspond to the exact and approximate breeding states, $P_{n,r_{\text{\tiny{opt}}}}^{\text{\tiny{Hom}}}(p)$ and $\tilde{P}_n^{\text{\tiny{Hom}}}(p)$, respectively.}
    \label{fig:FandDkl}
\end{figure}

We confirm the high fidelity between $\phi_{n,\delta}(q_1|p)$ and its exact breeding counter-part in $ \tilde{\phi}_{n,r_{\text{\tiny{opt}}},\tilde{\delta}}(q_1|p)$, for a few $p = \tilde{p}_l$, by numerically calculating 
\begin{align}
    F_{p}(r_{\text{\tiny{opt}}}) = \left| \int dq_1 ~  \tilde{\phi}^{*}_{n,r_{\text{\tiny{opt}}},\tilde{\delta}}(q_1|p) \phi_{n, \delta}(q_1|p)\right|^2.
\end{align}
Additionally, we show that the difference between $P_n^{\text{\tiny{Hom}}}(p)$ and $\tilde{P}_{n,r_{\text{\tiny{opt}}}}^{\text{\tiny{Hom}}}(p)$ is vanishing with increasing $r_{\text{\tiny{opt}}}$ by calculating the Kullback–Leibler divergence, $D_{\text{\tiny{KL}}}(r_{\text{\tiny{opt}}})$, between the two distributions. From the results in Fig. \ref{fig:FandDkl}, we can conclude that approximate state breeding approaches exact state breeding, as is expected given the high fidelity between the initial SCSSs and GPS states.

\section{Breeding with postselection Success Probability} \label{app:HomodyneProbability}
The total success probability when breeding GPS states with postselection is the product of the GPS state generation probability squared, the homodyne measurement success probability, and the damping operation success probability if applicable. 
That is,
\begin{align}
    P^{\text{\tiny{Total}}}_n = \left( P^{\text{\tiny{GPS}}}_{n}(r_{\text{\tiny{max}}}) \right)^2 \cdot P_n^{\text{\tiny{Sum}}} \cdot P_n^{\text{\tiny{Damp}}}(r_{\text{\tiny{d}}}^{\text{\tiny{opt}}}).
\end{align}
$P^{\text{\tiny{GPS}}}_{n}(r_{\text{\tiny{max}}})$ is the GPS success probability for a given $n$ when the input squeezed vacuum states have $r_{\text{max}}(n) = \cosh^{-1}\left(1+2 n\right)/2$,
\begin{align}
    P^{\text{\tiny{GPS}}}_{n}(r_{\text{\tiny{max}}}) & = \frac{2^{-n} n^n (1+2 n)^{-1/2-n} \Gamma(1+2 n)}{n!^2}.
\end{align}
Before calculating the homodyne measurement success probability, we must first determine what measurement outcomes constitute a success. In Appendix \ref{app:ExactStateBreeding}, we found that with exact state breeding, postselection on the outcomes $\tilde{p}_l$ in Eq. \eqref{eq:SSCBreedingConditions} heralded the state $\tilde{\phi}_{r}(q_1)$, in Eq. \eqref{eq:IdealBreedingState}. For each approximate breeding output state $\phi_{n}(q_1|p)$ we define a corresponding reference state $\tilde{\phi}_{r_{\text{\tiny{opt}}}}(q_1)$. We then find the measurement outcomes that maximize the fidelity between the two states on the intervals $ \tilde{p}_l - \sqrt{\pi}/2 \leq p \leq \tilde{p}_l + \sqrt{\pi}/2$ for $l \in \{l_{\text{\tiny{min}}},\ldots,l_{\text{\tiny{max}}}\}$ and collect them in the vector $\boldsymbol{p}$. $l_{\text{\tiny{max(min)}}}$ are chosen such that $\tilde{P}_n^{\text{\tiny{Hom}}}(\tilde{p}_{l_{\text{\tiny{max(min)}}}}) \geq 1 \%$. We then define the optimal measurement outcome to be 
\begin{align}
    p_{\text{\tiny{opt}}} & = \argmax_{\boldsymbol{p}} ~F_{n} (p)\label{eq:pOpt},\\
    F_{n} (p)  & = \left| \int dq_1~  \tilde{\phi}^{*}_{r_{\text{\tiny{opt}}}}(q_1) \phi_{n}(q_1|p) \right|^2
\end{align}
Next, fidelity maximization between $\phi_{n}(q_1|p)$ and $\phi_{n}(q_1|p_{\text{\tiny{opt}}})$ is performed on the same intervals, with the results being $\boldsymbol{p}'$. This vector is filtered down to contain only the outcomes that herald states with fidelity greater than $0.999$ to $\phi_{n}(q_1|p_{\text{\tiny{opt}}})$. Note that $p_{\text{\tiny{opt}}}$ is included in $\boldsymbol{p}'$. A homodyne measurement result of one of the elements of $\boldsymbol{p}'$ is considered to be a success, heralding the state $\phi_{n}(q_1|p_{\text{\tiny{opt}}})$. 

\begin{figure*}
    \centering
    \includegraphics[width=0.7\paperwidth]{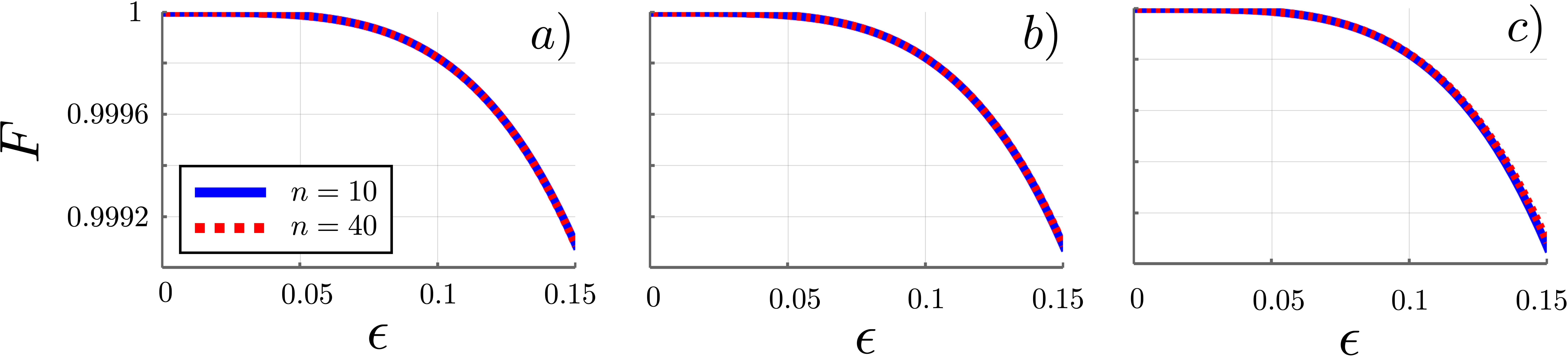}
    \caption{Fidelity between $\tilde{\phi}_{n,r_\text{\tiny{opt}}}(q_1|p)$ and $\tilde{\rho}_{n,r_\text{\tiny{opt}}}\left(q_1,q_1',\epsilon|p\right)$ as a function of $\epsilon$ for $n \in \{10,40\}$ and  (a) $p = 0$, (b) $p = \sqrt{\pi}$, and (c) $ p =2 \sqrt{\pi}$. }
    \label{fig:HomCheck1}
\end{figure*}

\begin{figure}
    \centering
    \includegraphics[width=0.75\linewidth]{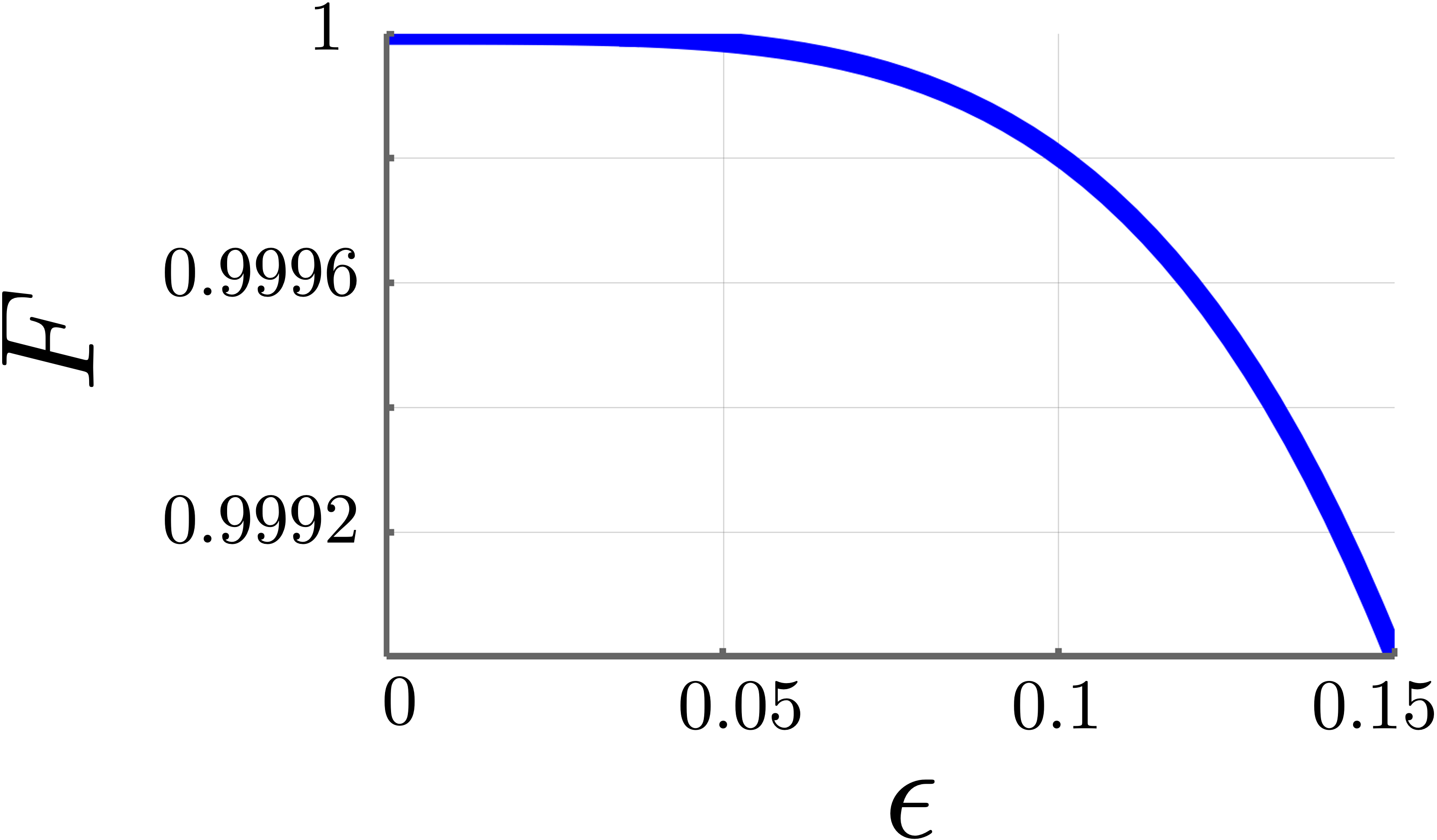}
    \caption{Fidelity between $\phi_{n = 8}(q_1|p=0)$ and $\rho_{n=8}\left(q_1,q_1',\epsilon|p=0\right)$ as a function of $\epsilon$.}
    \label{fig:HomCheck2}
\end{figure}

For a postselected homodyne measurement, we must take into account the finite-resolution of the measurement. Strictly speaking, a finite-resolution postselected homodyne measurement yields a mixed state, as shown in Eq.\eqref{eq:rhocon}. While the mixed state resulting from approximate breeding with finite-resolution homodyne detection can be derived, the subsequent corrective displacements and fidelity calculations are prohibitively difficult. To avoid such cumbersome calculations, we approximate the action of the operator in Eq.\eqref{eq:rhocon} by defining an acceptance window of width $2 \epsilon_i$ around each element of $\boldsymbol{p}'$, $p'_i$ such that $F \geq 0.999$ between $\phi_{n}(q_1|p_i)$ and $\phi_{n}(q_1|p_i \pm \epsilon_i)$. In all cases, we found $\epsilon_i \leq 0.12$.  The total homodyne postselection probability for heralding $\phi_{n}(q_1|p_{\text{\tiny{opt}}})$ is then calculated as
\begin{align}
    P_n^{\text{\tiny{Sum}}} & =\sum_{i=1}^{\text{\tiny{length}}(\boldsymbol{p}')} \int_{p'_i- \epsilon_i}^{p'_i+ \epsilon_i} dp ~ P_n^{\text{\tiny{Hom}}}(p).
    \label{eq:Psum}
\end{align} 
The validity of our approximation that $\rho_n(q_1, q'_1,\epsilon_i|p'_i)$ is well represented by $\phi_{n}(q_1|p_i)$ is demonstrated first by calculating the fidelity for exact breeding between $\tilde{\phi}_{n,r_\text{\tiny{opt}}}(q_1|p)$ and $\tilde{\rho}_{n,r_\text{\tiny{opt}}}\left(q_1,q_1',\epsilon|p\right)$ as a function of $\epsilon$ for $n \in \{10,40\}$ and $p \in \{0, \sqrt{\pi}, 2 \sqrt{\pi}\}$. As can be seen in Fig. \ref{fig:HomCheck1}, for $\epsilon \leq 0.15$, the fidelity between $\tilde{\phi}_{n,r_\text{\tiny{opt}}}(q_1|p)$ and $\tilde{\rho}_{n,r_\text{\tiny{opt}}}\left(q_1,q_1',\epsilon|p\right)$ is above $0.999$. For approximate breeding, we calculate the fidelity between $\phi_{n = 8}(q_1|p=0)$ and $\rho_{n=8}\left(q_1,q_1',\epsilon|p=0\right)$ and plot it as a function of $\epsilon$ in Fig. \ref{fig:HomCheck2}. Similarly, for $\epsilon \leq 0.15$, $\phi_{n = 8}(q_1|p=0)$ and $\rho_{n=8}\left(q_1,q_1',\epsilon|p=0\right)$ are nearly indistinguishable. Given that approximate breeding approaches exact breeding as $n$ increases, as established in Appendix \ref{app:ExactStateBreeding}, it is reasonable to assume that $F \geq 0.999$ between $\phi_{n}(q_1|p)$ and $\rho_{n}\left(q_1,q_1',\epsilon|p\right)$ when $\epsilon \leq 0.15$ will hold for higher $n$ and $p$. Therefore, $\rho_n(q_1, q'_1,\epsilon_i|p'_i)$ is well approximated by $\phi_{n}(q_1|p_i)$ with our use of $\epsilon_i \leq 0.12$.

The success probability of the damping operation in Ref. \cite{takase2023gottesman}, $P_n^{\text{\tiny{Damp}}}(r_{\text{\tiny{d}}}^{\text{\tiny{opt}}})$, is 
\begin{align}
    P_n^{\text{\tiny{Damp}}}(r_{\text{\tiny{d}}}^{\text{\tiny{opt}}}) & = \int dq_1 ~ \left| \mathcal{D}_{r_{\text{\tiny{d}}}^{\text{\tiny{opt}}}}\left[\Psi\right](q_1) \right|^2 \\
    \mathcal{D}_{r_{\text{\tiny{d}}}^{\text{\tiny{opt}}}}\left[\Psi\right](q_1) & = \frac{\phi_{n}(q_1|p_{\text{\tiny{opt}}})}{e^{\frac{1-\tanh\left(r_{\text{\tiny{d}}}^{\text{\tiny{opt}}}\right)}{4}q_1^2}} \sqrt{ \sech \left(r_{\text{\tiny{d}}}^{\text{\tiny{opt}}}\right)}.
\end{align}

\bibliography{PRL_Submission}

\begin{thebibliography}{41}
\bibitem [{GKP()}]{GKPQubit}%
  \BibitemOpen
  \href@noop {} {}\bibinfo {note} {The finite-squeezed GKP qubit is made of the states \begin{align} \ket{\tilde{k}_{\xi}} & = \nonumber \\ & \mathcal{N}_{k,\xi}\left[\sum_{t=-\infty}^{\infty} e^{-\frac{e^{-2\xi}}{2}((2t+k)\sqrt{\pi})^2 } \hat{X}\Big((2t+k)\sqrt{ \pi}\Big)\right]\ket{S_\xi}, \nonumber \end{align} where $\mathcal{N}_{k,\xi}$ is a normalization constant, $k \in \{0,1\}$, $\hat{X}(q) = \hat{D}(q/\sqrt{2})$, and $\ket{S_\xi} = \hat{S}(\xi)\ket{0} $. The conversion of $\xi$ to dB is $\xi_{\text{dB}} = 10\log_{10}(e^{2 \xi})$}\BibitemShut {NoStop}%
\end{thebibliography}


\end{document}